\documentclass{ws-rv961x669}
\usepackage{ws-rv-van}     
\usepackage{ws-rv-thm}     
\usepackage{subfigure}     
\makeindex

\begin{document}

\chapter[Using World Scientific's Review Volume Document Style]{Can one determine the neutrino mass by electron capture?}

\author{Amand Faessler\footnote{e-mail: faessler@uni-tuebingen.de}}

\address{Institute of Theoretical Physics,
University of Tuebingen, Germany, \\  \vspace{0.5 cm} \large{Dedicated to the Memory of Walter Greiner, 1935 - 2016.}\\ }
\begin{abstract}
There are three different methods used to search to determine the neutrino mass:\begin{enumerate} 
\item The electron antineutrino mass can probably best be determined by the Triton \\ decay. \item The neutrinoless Double Beta Decay yields information, if the neutrino is a \\ Dirac or a Majorana particle. It can also determine the Majorana neutrino mass. \item Electron capture of an atomic bound electron by a proton in a nucleus \\ $e\  +\  p \ \rightarrow \ n\ + \ \nu \ $  can give the mass of the electron neutrino. \end{enumerate}
This contribution discusses the possibility to determine the electron neutrino mass by electron capture. One expects the largest influence of the neutrino mass on this decay for a small Q = 2.8 keV $^{163}_{67}Ho\ + \ e \ \rightarrow \ ^{163}_{66}Dy \ + \ \nu$. The energy of the Q value is distributed to the emitted neutrino and the excitation of the Dy atom. Thus the energy difference between the Q value and the upper end of the spectrum is the electron neutrino mass. The excitation spectrum of Dy is calculate by one-, two- and three electron hole excitations, and by the shake-off process. The electron wave functions are calculated selfconsistently by the Dirac-Hartree-Fock approach for the bound and the continuum states. To extract the neutrino mass from the spectrum one must adjust simultaneously the neutrino mass, the Q value, the position,the relative strength and  the width of the highest resonance. This fit is only possible, if the background is reduced relative to the present situation. The analysis presented here shows, that the determination of the electron neutrino mass by electron capture is difficult, but seems not to be impossible. 

\end{abstract}


\body

\section{Introduction}\label{Intro}

Neutrinos are one of the least known particles, although they play a central role in weak interactions. We do not know the mass of the three neutrinos, electron, muon and tauon neutrinos (and antineutrinos). We do not know their hierarchy. Which one is the heaviest? The nature of the neutrinos, Dirac (neutrino different from the antineutrino) or Majorana (neutrino identical with the corresponding antineutrino) is still unknown. At the moment one searches with three different methods for the neutrino mass. The Tritium decay \cite{Otten, Drexlin} can measure the antineutrino mass, while electron capture can determine the the neutrino mass \cite{Fae1,de,Fae3,Robertson,Fae4,de2}. The neutrinoless Double Beta Decay can differentiate beween Diray and Majorana neutrinos and can give the Majorana neutrino mass \cite{Fae5,Sim1,Vo1,Av1}.
\\   \linebreak
Here I summarizes the work on the determination of the electron neutrino mass by electron capture in $^{163}_{67}Ho$ of the Tuebingen group and coworkers \cite{Fae3,Fae7,Fae8,Fae3hole,Fae1-2hole}.
Elctron capture is most sensitive for the neutrino mass for a small Q-value: For the capture 
\begin{eqnarray}
^{163}Ho + e  \rightarrow  ^{163}Dy  + \bar{\nu}
\label{capture}
\end{eqnarray}
 the value obtained by a Penning trap measurement by Eliseev et al. \cite{Eliseev} (see also \cite{Ra,Lo2,Blaum,Ra2}) is: 
\begin{eqnarray} 
Q = 2.833 (30_{stat}) (15_{sys}) \hspace{0.1 cm} keV  
\label{Qvalue} 
\end{eqnarray}
This means, that electron capture in $^{163}Ho$ is only possible from electron 
orbits bound less than Q = 2.8 keV. So capture is only possible from $3s_{1/2}, M1$ 
with a binding energy 2.128 keV and less bound electron states. In addition the electron orbits in Ho, from 
which an electron can be captured must have an overlap with the nucleus to react 
with a proton. This is only the case for $(n>2)s_{1/2}$ and the lower relativistic amplitude 
of the $(n>2)p_{1/2}$ states. Thus a relativistic treatment of the electron wave 
functions is essential. 
The energy of the Q value after capture is distributed over the excitation energy 
of the Dy atom (one-, two-, three-hole, ... excitations), the emitted neutrino and 
the recoil of the Ho nucleus. The recoil of the Ho  nucleus can be neglected and we 
have to consider the distribution of Q to the excitation of Dy and to the neutrino only. 
The excited Dy decays by X-rays and Auger electron emissions. The sum of these energies 
can be measured by a bolometer. The maximum bolometer energy yields then the 
minimum neutrino energy, which is the electron neutrino rest mass. Thus the neutrino 
mass can be determined by the difference of the Q-value and the upper end of the 
bolometer spectrum. 
The excitation energy of the Dy* is calculated in the sudden approximation. The 
excitation of the Dy* is then given by the overlap with Ho* with an electron hole 
in the orbit of the capture with the different excited configurations $|D>$  
in the daughter Dy*. Lets estimate before we start the detailed calculations roughly an upper 
limit for the probability of exciting an electron in Dy* into the continuum (shake-off). 
We start with the normalisation of the many electron one-hole state in Ho, 
from which the electron is captured and use the completeness relation. 
\begin{eqnarray} 
1\  = \  <P,\  (n, \ell_{1/2})^{-1},\ Ho\ |P,\ (n, \ell_{1/2})^{-1},\ Ho> \ =  \nonumber  \hspace{8cm} \\
\sum_{D,Dy} \ <P, \ (n,\ell_{1/2})^{-1},\ Ho|D,\ Dy><L,\ Dy|P, (n, \ell_{1/2})^{-1} , \ Ho> \ =
 \nonumber \hspace{7cm}  
\end{eqnarray}
\begin{eqnarray}
|<P, \ (n,\ell_{1/2})^{-1}, Ho|P',\ Dy>|^2 +  
\sum_{D \neq P} |<P,\ (n,\ell_{1/2})^{-1},Ho|D,\ Dy>|^2 
\label{overlapA}
\end{eqnarray}
"P" stands here for configurations in the Ho parent  (Not to be mixed up with the large relativistic electron continuum $P_E(r)$ or bound state $P_b(r)$ amplitudes, which normally have a l.c. index.) and 
"D" for a configuration in the Dy* daughter. The star indicates excited states. 
The first term in the last line of (\ref{overlapA}) is a diagonal overlap squared and thus the probability, that the electron Dy configuration $|P'>$  with 66 electrons has a hole for the same quantum numbers from which the electron is captured in Ho (contriutions to one-hole excitations). The prime reminds, that DHF electron orbits of Dy must be used.  To get rid of the many electrons, which are practically not involved in the capture process and for which the wave functions have an overlap $<nlj,Ho|nlj,Dy> \approx 0.999$ between Ho and Dy, we use the Vatai approximation \cite{Vatai1, Vatai2} and put these overlaps for this first estimate to unity. for the final results in section 2 to 5 we calculate these overlaps correctly and do not use the Vatai approximation \cite{Vatai1, Vatai2}.  For the first term in the last line of(\ref{overlapA}), which is the probability, that the configuration 
$|P,(n,\ell_{1/2})^{-1},Ho>$ is "not changing" in Dy, one has to consider 
the (2j +1) = 2 magnetic substates for the $ns_{1/2}$ and $np_{1/2}$ states, from which electrons can be captured. This yields for the first term in the last line of eqs. (\ref{overlapA}) 
an exponent 2(2j+1) = 4. The last term of (\ref{overlapA}) contains the non-diagonal overlaps between Ho and Dy configurations 
and is the probability for shake-up and for the shake-off excitations  in Dy* configurations $|D>$  
(different from the one-hole configurations  $|P'>$,  contained in the first term of the last line of (\ref{overlapA})) by moving electrons up into free bound states (shake-up) or into the continuum (shake-off). 

The diagonal overlaps (first term of the last line of (\ref{overlapA}) are very close to unity and are typically 
$<nlj,Ho|nlj,Dy> = 0.999$, since the electron wave functions do  only change minimal from $_{67}Ho$ to $_{66}Dy$. To get a  feeling how large the shake-off probability can roughly be, we use for a first estimate for the diagonal overlap as a measure for the deviation from the Vatai approximation $<P, (nlj)^{-1},Ho|P', (nlj)^{-1},Dy> = 0.999$ ( See table \ref{over} ). The probability to excite Dy* into a shake-off state  is then 
less than one minus the probability, that in Dy the configuration $|P'>$ is not changed. The probability to excite two-, three- and many-hole states (shake-up) and to excite continuum states (shake-off) is one minus the probability, that the Configutation $|P>$ of the excited Ho is not changed in Dy*. 

\begin{eqnarray}
Prob_{shake-off} \le [1.0 - < (n, \ell,j)^{-1},\ Ho|(n,\ell,j)^{-1}, \ Dy>^{2\cdot (2j+1)}] \nonumber \\ 
\approx [1.0 -0.999^4] = 0.004 \equiv 0.4\  \% 
\label{VataiA}
\end{eqnarray}
Thus the shake-off process should be less than $0.4 \  \%$ relative to the one-hole probability.  
If perturbaton theory is used as in ref. \cite{de2} to calculate this overlap of Ho and Dy states, one easely has a $ 10 \% $ error or more: $ < (n, \ell,j)^{-1},\ Ho|(n,\ell,j)^{-1}, \ Dy>\ \approx \ 0.9$:  
\begin{eqnarray}  
Prob_{shake-off} \le [1.0 - 0.90^{4}]  \approx 0.34 \equiv \ 34\ \%;  \nonumber \\  34 \ \% \  of \ the \  1-hole\ states. 
\label{VataiB}
\end{eqnarray}

Thus with such an error in the overlap the shake of probability can be  about $34 \%$ of the one hole states and by about a factor 85 larger than a more correct estimate. This result requires very good electron wave functions for a reliable description of the capture process. 
DeRujula and Lusignoli \cite{de2} use non-relativistic pure Coulomb wave functions for the electrons in Ho, which do not allow capture from $np_{1/2}$. The effect of the partial screening of the nuclear Coulomb potential by the electron cloude  is considered by effective charges. Since in 
reference \cite{de2} the perturbation approach is used to calculate the Dy* wave functions, this requires orthonormal basis functions. 
Thus ref. \cite{de2} uses for each capture process, e. g. 
from $3s_{1/2}$, the same screening factor for all single electron orbits needed in the perturbation approach for this special capture process. This means the screening is paradoxically the same for inner and for outer electron shells for a specific capture process. 
Starting from these "bad" wave functions for Ho ref. \cite{de2} calculates the bound and continuum wave functions for Dy* with the perturbation:
\begin{eqnarray}
  H'(r) =\ +1/r\ -\ \int d^3r_1 \ |\varphi_{3s/4s}(\vec{r}_1)|^2/|\vec{r} - \vec{r}_1|.    
\label{pertur}
\end{eqnarray}
The first term takes into account, that in Dy one has one proton less compared to Ho. The minus sign of the p-e interaction yields the plus sign in (\ref{pertur}). The second term originates from the electron hole in the $3s_{1/2}$ or the $4s_{1/2}$ state, which are the only hole states considered in ref. \cite{de2}.
Equation (\ref{overlapA}) serves as a lever to enlarge a small change in the overlap into a large change for the shake-up and shake-off probabilities. A $ 10\ \%$ change in the overlap can produce an enlargement of almost a factor 100 for the shake-up and shake-off probabilities. Thus one needs very good electron wave functions, else the 
capture probabilities are not reliable. 

We use here selfconsistent Dirac-Hartree-Fock wave functions caculated specifically for Ho* and Dy* \cite{Grant, Desclaux, Ankudinov}. The continuum electron wave functions in Yb* are determined \cite{Salvat}  in the selfconsistent Z-1 = 65 electron potential orthogonal to all 65 bound electrons in Yb. 

The improvements relative to Inteman and Polok \cite{Inte} and to DeRujula and Lusignoli \cite{de2} are: 
\begin{itemize}  
\item  
The sudden approximation \cite{Fae3,Fae4, Fae7, Fae8, Fae3hole,Fae1-2hole} with selfconsistent Dirac-Hartree-Fock (DHF) wave functions for the Dy atom  is used to determine the electron capture probability and not the less reliable first order perturbation theory \cite{Inte,de2}.
\item  We do not use for $^{163}_{67}Ho$ non-relativistic screened Coulomb functions \cite{Inte,de2}, but calculate all electron  wave functions by the relativistic, selfconsistent Dirac-Hartree-Fock approach \cite{Ankudinov, Grant, Desclaux}  
with full antisymmetrization. Among many other advantages the electron orbitals are in this way all orthogonal.
\item The wave function of the bound electrons in Dysprosium are again determined selfconsistent and relativistic by Dirac-Hartree-Fock \cite{Ankudinov, Grant, Desclaux} even allowing for 3s and 4s hole states. In ref. 
\cite{Inte} and  in ref. \cite{de2} the electron orbitals for the daughter Dy are calculated in first order perturbation theory (\ref{pertur}) from the Coulomb functions of Ho. 
\item The continuum s-wave electron for shake-off in Dy is calculated relativistically in the selfconsistent DHF potential  of the selfconsistent 65 electron core, under the condition, that the continuum s-orbitals are orthogonal the the bound s-orbits in Dy. 
\item Numerical stability is tested carefully. For the continuum electron wave functions in Dy for the radial coordinate 250 up to more than 700 mesh points were used depending on the energy. The integration over the continuum electron energies for the shake-off electron are performed from 0 to Q = 2.8 [keV] with 417 mesh points. Integrations for the norms, the overlaps and the integration over the shake-off in the continuum were done in parallel with the Trapez rule (error $\propto$ second derivative), the Kepler-Simpson rule (error     
$\propto$ fourth derivative), the Bode-Boole rule (error $\propto$ sixth derivative) and the Weddle rule. From the points of stability and accuracy the Bode-Boole's rule  turned out to be the most reliable. All the calculations  were done in double precision. 
\item The DHF overlaps of $<3s,Ho|3s,Dy>\ = \ 0.99940$  and $<4s,Ho|4s,Dy>\  =\  0.99909$ (See table \ref{over}.) limit in the Vatai approximation \cite{Vatai1, Vatai2} the 2-hole probability including the shake-off process, which requires a second hole, to $ 0.24 \% $ and $0.36 \%$ of the 1-hole excitations. An error of $ 10 \%$ in calculating the single orbital overlaps between  Ho and Dy due to first order perturbation theory \cite{Inte,de2} estimated again with the Vatai approximation can increase the shake-off probability by two orders of magnitude. Eq. (\ref{overlapA}) serves as lever to produce from a small uncertainty  of the single electron overlaps a large increase of the shake-off probability. If one does not use the Vatai approximation and puts all electron orbital overlaps of Ho with Dy  to 0.999,  the definite upper limit (including 1- and  2-hole and shake -off excitations)  for shake-off is 12 \% relative to the 1-hole states. The norm gives without Vatai no restriction for the shake-off with an error of 10 \% for the  $<n,\ell,j,Ho|n,\ell,j,Dy>$ single electron overlaps 
\item In this work the different 1-hole, 2-hole and shake-off contributions are taken from the theory without adjusting them in different ways to fit the experiment. In ref. \cite{de2} the authors write: 
\newline " Our estimate of the height of the N1(4s)O1(5s) shakeup peak is a factor $\approx 2.5$ too low. It is possible to correct in similarly moderate ways the other contributions such as to agree with the data." 
\end{itemize}
\vspace{1cm}
\begin{table}[ht]
\tbl{Overlaps of the 3s and 4s wave functions of Ho with Dy. DHF (Dirac-Hartree-Fock) is the result of the selfconsistent, relativistic calculation of this work. $1- <DHF>^4$ gives the probability, that Dy* is excited beyond the one-hole states  including shake-up and shake-off. Thus the the shake-off must be less than $0.4 \%$ of the one-hole excitation probabilities in disagreement with reference \cite{de2}. "Coulomb" shows the overlaps with  non-relativistic Coulomb wave functions and effective charges $Z(3s)_{eff} = 54.9$  and $Z(4s)_{eff} = 43.2$ used by DeRujula and Lusignoli \cite{de2}. Here the Coulomb wave functions are calculated in Ho and Dy separatly. In refs. \cite{de2,Inte} the Coulomb wave functions in Dy are calculated with perturbation theory from Ho using expression (\ref{pertur}) as perturbation.  The diagonal overlaps show the right oder of magnitude, but the non-diagonal elements of the overlap are completely wrong. Even the orthogonality of 3s and 4s in the same atom is not fullfilled. \vspace{0.3cm}} 
{\begin{tabular}{|c |r|r|r|r|} \hline
$  -------  $& $ DHF$ &$ 1 - <DHF>^4 $&$ Coulomb $&$ 1 - <Coul>^4$\\ \hline \hline
$<3s, Ho| 3s, Dy>$& 0.99940 & 0.00239 & 0.99932 & 0.00271 \\ \hline
$<4s, Ho| 4s, Dy>$& 0.99909 & 0.00363 & 0.99848 & 0.00607 \\ \hline
$<3s, Ho| 4s, Dy>$&-0.01982 &   ---   & 0.56828 &   ---   \\ \hline
$<4s, Ho| 3s, Dy>$& 0.02067 &   ---   & 0.56817 &   ---   \\ \hline
$<3s, Ho| 4s, Ho>$& 0.0     &   ---   & 0.56857 &   ---   \\ \hline
$<3s, Dy| 4s, Dy>$& 0.0     &   ---   & 0.56952 &   ---   \\ \hline \hline
\end{tabular}}
\label{over}
\end{table}
\vspace{1cm}
\pagebreak

\section{Description of Electron Capture}\label{Capture}
The spectrum of the the decay of the excited $^{163} Dy^*$  
after electron capture in $^{163} Ho$ refs. \cite{de, de2} and \cite{Fae3} 
assuming Lorentzian shapes for the decay resonances are: 
\begin{eqnarray} 
 \frac{d\Gamma}{dE_c} \propto \sum_{i = 1,...N_\nu}(Q - E_c)
\cdot U_{e,i}^2\cdot\sqrt{(Q-E_c)^2 -m_{\nu,i}^2}* \hspace{14cm} \nonumber \\
(\sum_{f=f'} \lambda_{0}B_f \frac{\Gamma_{f'}}{2\pi} 
\frac{1}{(E_c - E_{f'})^2 +\Gamma_{f'}^2/4} + \hspace{16cm} \nonumber \\
\sum_{f=f';p'<F;q'_b >F} \lambda_{0}B_{f,p'<F;q'_b>F} \frac{\Gamma_{f',p'}}{2\pi} 
\frac{1}{(E_c - E_{f',p'})^2 +\Gamma_{f',p'}^2/4}+   
\hspace{14cm} \nonumber  \\ 
 \int dk_{q'} \lambda_{0}B_{f,p'<F;q'_c>0} \frac{\Gamma_{f',p',q'}}{2\pi} 
\frac{1}{(E_c - E_{f',p',q'})^2 +\Gamma_{f',p',q'}^2/4} ) \hspace{2cm} (7) \hspace{12cm} \nonumber \\ \label{decay}
\end{eqnarray}
The factor in front of the bracket is the same as for the single beta decay. It reflects the phase space of the emitted neutrino 
 (\ref{capture}). 
The three expressions in the bracket of eq. (\ref{decay}) describe the decay of the 1-hole f' excitations in Dy*, of the  2-hole excitations  f', p' with  shake-up of p' to q' into a bound orbit and of 2-holes f', p' states with one electron p' moved to q' into the continuum (shake-off). The last term has as the others a dimensionless strength $dk_{q'} \cdot B_{f',p'<F,q'>o}$. The transformation to an integral over the energy  yields the factor (\ref{energy}).  Different neutrino mass eigenstates $|i>$ are mixed into the electron neutrino  flavor states by $U_{e,i}^2$.   
The Q-value is given by eq. (\ref{Qvalue}) from the ECHo collaboration \cite{Blaum, 
Anderson, Gatti, Ra, Audi}.
The excitation energy of the final Dy* is $E_c$. The energy $Q\ -\ E_c$ is carried away by the neutrino. $B_f $, $B_{f,p'<F;q'_b>F}$ and $B_{f,p'<F;q'_c>0}$  are the overlap and exchange corrections for the 1-hole, the bound 2-hole and the shake-off 2-hole states. $\lambda_{0}$ contains 
the nuclear matrix element squared \cite{bam}. Eq. (\ref{decay}) yields the spectrum in arbitrary units and must be normalized at one energy to the data. We use the N1, $4s_{1/2}$ peak for the normalisation.   
$ E_{f'}$, $E_{f',p'}$ and $E_{f',p';q'>0}$  are the 1-hole, the 2-hole shake-up and the 2-hole shake-off excitation energies in Dysprosium (see tables \ref{binding} and \ref{2-binding}). 
$ \Gamma_{f'}$, $\Gamma_{f',p'}$ and $\Gamma_{f',p';q'>0}$ are the widths of the one- and two-hole states and the two-hole states with shake-off  in Dysprosium \cite{Fae7,Fae8, Fae3hole, Fae1-2hole}. 
%
%
%
\begin{table}[ht]
\tbl{Important electron binding energies in $^{163}_{67}Ho$ \cite{Williams}. Electrons can only be captured from orbitals overlapping with the nucleus. This restricts capture to $ns_{1/2}$ and in a relativistic treatment to the lower amplitude of $np_{1/2}$. Energy conservation requires, that the Q-value Q = 2.8 keV must be larger than the binding energy of the electron captured in Ho. This is the case for $3s_{1/2}$ and higher levels. \vspace{0.5 cm}} 
{\begin{tabular}{|r|c|c|} \hline
$      n\ell_{j}  $&     $   Notation\  for\  Electron\  Orbit   $ &    $   Binding\  Energy\  E_b [keV]  $\\ \hline \hline
$  1s_{1/2}$&$ K1$& 55.618  \\ \hline
$  2s_{1/2}$&$ L1$&  9.394  \\ \hline
$  2p_{1/2}$&$ L2$&  8.918  \\ \hline
$  3s_{1/2}$&$ M1$&  2.128  \\ \hline
$  3p_{1/2}$&$ M2$&  1.923  \\ \hline
$  4s_{1/2}$&$ N1$&  0.4324 \\ \hline 
$  4p_{1/2}$&$ N2$&  0.3435\\ \hline \hline
\end{tabular}}
\label{binding}
\end{table}
%
%
%
%
\begin{table}[ht]
\tbl{Electron binding energies and width of two-hole states in $^{163}_{66}Dy$, which contribute to s-wave shake-off. Energy conservation requires, that the Q-value Q = 2.8 keV must be larger than the two-hole binding energy plus the energy of the electron in the continuum. The shake-off contributions for the 2-hole states start at the 2-hole binding energy in the bolometer spectrum as a function of $E_c$. The width includes only the contribution from the decay of the 2-hole states, but not the escape width of the continuum electron.\vspace{0.5cm}} 
{\begin{tabular}{|c |c|c|} \hline
$ n_1\ell_{j,1}, n_2\ell_{j,2} \ $& $ 2-hole \ Binding\ Energy \   E_b [keV]$&$ Width\ [keV]\ $\\ \hline \hline
$3s_{1/2}, 4s_{1/2}$& 2.4742& 0.0264 \\ \hline
$4s_{1/2}, 4s_{1/2}$& 0.8414& 0.0108 \\ \hline
$4s_{1/2}, 5s_{1/2}$& 0.4583& 0.0107 \\ \hline
$4s_{1/2}, 3p_{1/2}$& 2.2692& 0.0114 \\ \hline
$5s_{1/2}, 3p_{1/2}$& 1.8861& 0.0114 \\ \hline
$3s_{1/2}, 4p_{1/2}$& 2.3853& 0.0186 \\ \hline 
$4s_{1/2}, 4p_{1/2}$& 0.7525& 0.0107 \\ \hline
$5s_{1/2}, 4p_{1/2}$& 0.3776& 0.0106 \\ \hline \hline
\end{tabular}}
\label{2-binding}
\end{table}

The overlap and exchange corrections $B_f $, $B_{f,p'<F;q'_b>F}$ and $B_{f,p'<F;q'_c>0}$ are calculated in second quantisation. 
This yields automatically the antisymmetrisation and the exchange terms. To obtain these quantities we need:

\begin{eqnarray}
  |G> =  a_1^{\dagger} a_2^{\dagger} a_3^{\dagger}... a_Z^{\dagger} |0> \label{G}
\end{eqnarray}
The creation and anihilation operators of selfconsistent DHF electron orbits in Dy $a'^{\dagger}_i$ and $ a'_i$ are characterized by a prime. The corresponding operators in Ho are unprimed. 
\begin{eqnarray}
  |A'_{f'}> =  a'^{\dagger}_1 a'^{\dagger}_2... a'^{\dagger}_{f'-1}a'^{\dagger}_{f'+1}... 
	a'^{\dagger}_{Z} |0> \label{A}
\end{eqnarray}
The antisymmetrized two-hole state in Dy with shake-off is:
\begin{eqnarray}
  |A'_{p',f':q'>0}> =  a'^{\dagger}_1 a'^{\dagger}_2... a'^{\dagger}_{f'-1}a'^{\dagger}_{f'+1}... 
	a'^{\dagger}_{p'-1}a'^{\dagger}_{p'+1}...a'^{\dagger}_{Z}a'^{\dagger}_{q' > 0} |0> \label{A2}
\end{eqnarray}
The probability to form a two-hole shake-off state is proportional to:
\begin{eqnarray}
 P_{f',p';q'>0} =  |<A'_{f', p'; q'}|a_i|G>|^2  \label{Bffff}
\end{eqnarray} 
The relative shake-off probability normalized to the 3s one hole excitation is: 
\begin{eqnarray}
 B_{f',p';q'>0} = \frac{|\psi_{f}(R) <A'_{f',p':q'<0}|a_f|G>|^2}{|\psi_{3s1/2}(R)|^2} 
= P_{f',p';q'>0} \cdot \frac{|\psi_{f}(R)|^2}{|\psi_{3s1/2}(R)|^2}  
\label{Bff}
\end{eqnarray}
Normally the wave function of the captured Ho electron  is taken for the nuclear matrix element at the origin. Here we take this electron wave function at the nuclear radius. Due to the weight $r^2$ of the integration this is a better choice. 
\begin{eqnarray}
 P_{f',p';q'>0} = |<0|a'_{q'}a'_Z...a'_{p'+1}a'_{p'-1}...
a'_{f'+1}a'_{f'-1}...a'_{1'}\cdot a_f \cdot a^+_1...a^+_Z|0>|^2 = \hspace{1cm}  \nonumber \\
|<A'_{p',f'<F; q'>F}(2\ holes)|a_f|G>|^2 
\approx   |<q'_{>0}|p_{<F}>  \cdot \prod_{k=1..Z \neq f,p}<k'|k> |^2  \hspace{1cm}
\label{con}
\end{eqnarray}
To evaluate the probability for the shake-off process the excitations are restricted to s-waves.
\begin{eqnarray}
P_{f', p'} =  \sum_{q' > F} |<p_{<F, Ho}|q'_{>F, Dy}><q'_{>F, Dy}|p_{<F, Ho}>| \hspace{2cm} \nonumber \\ 
\cdot \prod_{k=k'<F_{Dy} \neq f,p}|<k'_{Dy}|k_{Ho}>|^2 \hspace{4cm} 
\label{F22}
\end{eqnarray}
\section{The Dy Continuum wave functions.}
\begin{figure}
\centerline{\includegraphics[width=11cm,angle=-90]{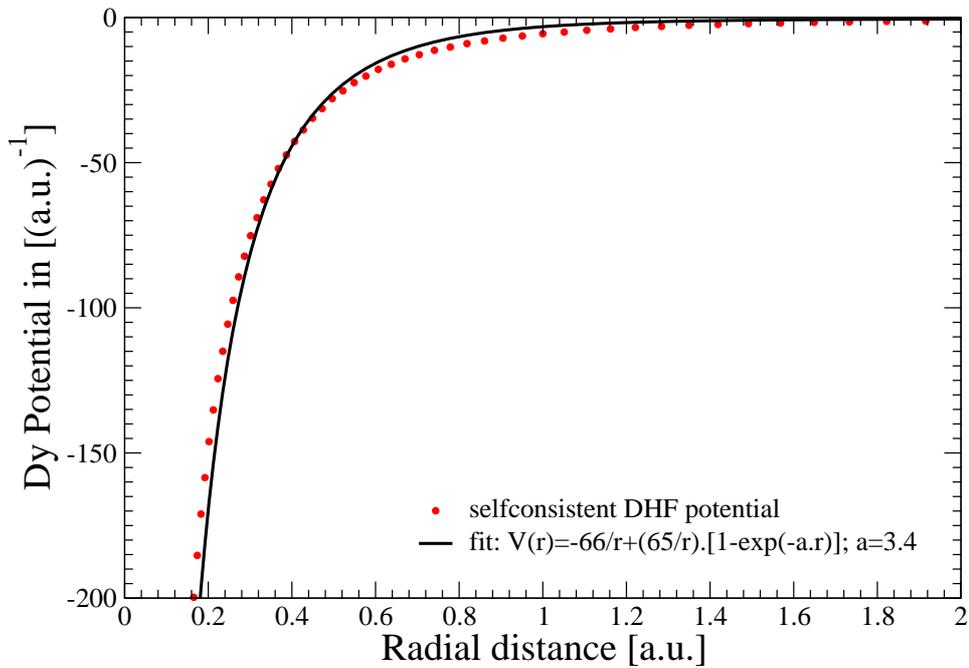}}
\caption{Selfconsistent DHF potential in Dy* with Z = 66 for a continuum electron (No 66) of 65 electrons 
(dimension: 1/length = atomic units) in the core and analytic approximation with a = 3.4 (See eq. \ref{pot3}). }
\label{Fig1Greic}
\end{figure}

\begin{figure}
\centerline{\includegraphics[width=11cm,angle=-90]{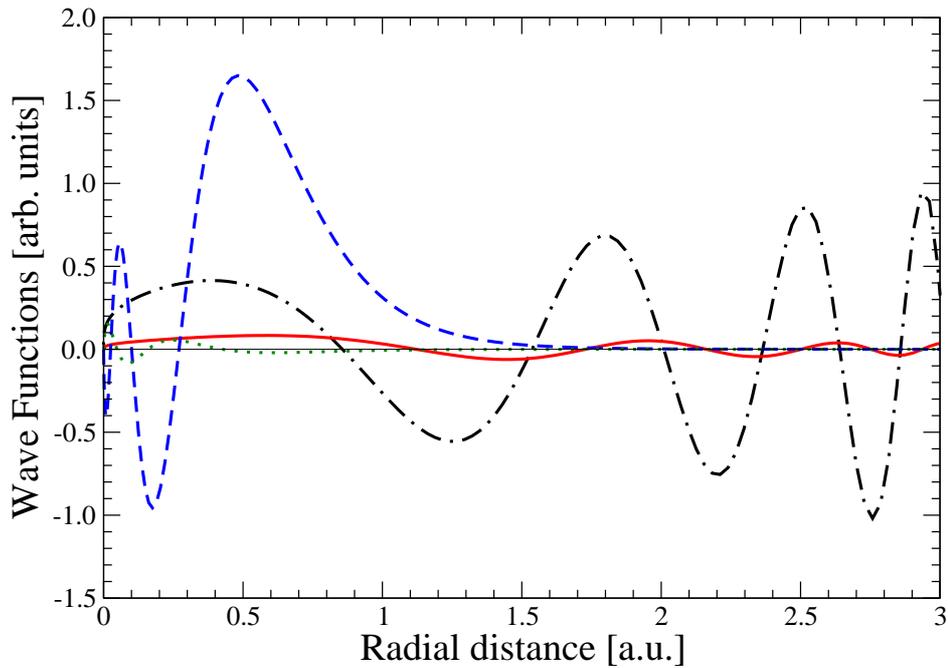}}
\caption{The electron wave function $P_{E}(r)$ (dashed-dot) and $Q_{E}(r)$ (solid) (\ref{PQ})  in the Dy continuum at 50 [Hartree] = 1.36 keV. The continuum wave functions $P_E$ and $Q_E$ are normalized to $\int  dr [P_{E}(r)^2 + Q_{E}(r)^2\ ] = \  2\pi \ \cdot  \delta (k - k')$. The 4s bound state in Ho is normalized to unity (dashed for$P_{4s}(r)$ and dotted for $Q_{4s}(r)$): $\int dr 
[P_{4s}(r)^2 + Q_{4s}(r)^2 ] = 1 $. The overlap $<(ns)_{-1},\ Ho|E, \ s,\ Dy> $ squared is proportional to the shake-off process as a function of energy. The continuum wave functions $P_E\  and\  Q_E$ are dimensionless, while the bound states $ P_{4s}(r)\  and\  Q_{4s}(r)$ are in atomic units $[(a.u.)^{-1/2}]$.}
\label{Fig2Greic}
\end{figure}

The Dy continuum wave functions for the shake-off require the following:
\begin{itemize}
\item The continuum electrons move in the Coulomb potential of the Dy nucleus $ V_{nucleus} = -66/r$ and of the selfconsistent cloude of 65 core electrons allowing for  the different empty states ( figure \ref{Fig1Greic}). 
\item The relativistic continuum electron wave functions have to be orthogonal to all bound orbits in Dy. We determine them  with the code of Salvat et al. \cite{Salvat} and Schmidt orthogonalization ( figure \ref{Fig2Greic}). 
\item These continuum functions \cite{Salvat} are normalized to delta functions in wave numbers $2 \ \pi \cdot \delta(k-k') $.  To integrate over the excitation energy in the continuum (\ref{decay}), the y have to be normalized 
to energy delta functions $\delta(E - E')$.
\end{itemize}
\begin{figure}
\centerline{\includegraphics[width=11cm,angle=-90]{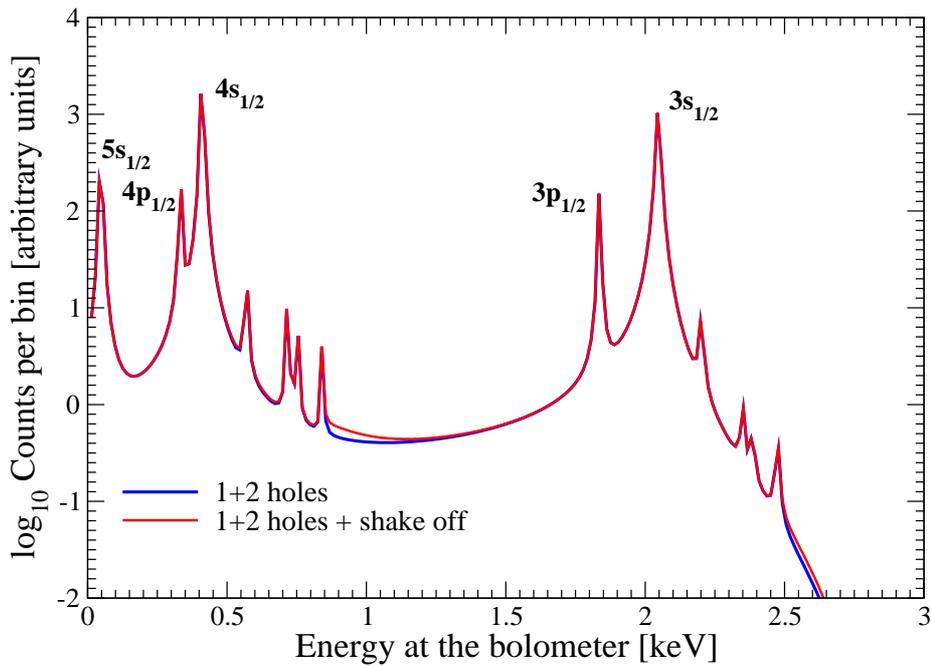}}
\caption{Theoretical results in arbitrary units of the sum of the one- and two-hole deexcitations compared to the sum of the one-, two-hole and the shake-off deexcitation as measured by the bolometer spectrum (\ref{decay}). The arbitrary units are adjusted to the experimental N1, $ 4s_{1/2} $ 1-hole peak (see figure 10). The nature of the one hole states are indicated. The two-hole peaks are by about two orders of magnitudes smaller than the one hole peaks. Shake-off can almost not been seen.}
\label{Fig3Greic}
\end{figure}
\begin{figure}
\centerline{\includegraphics[width=11cm,angle=-90]{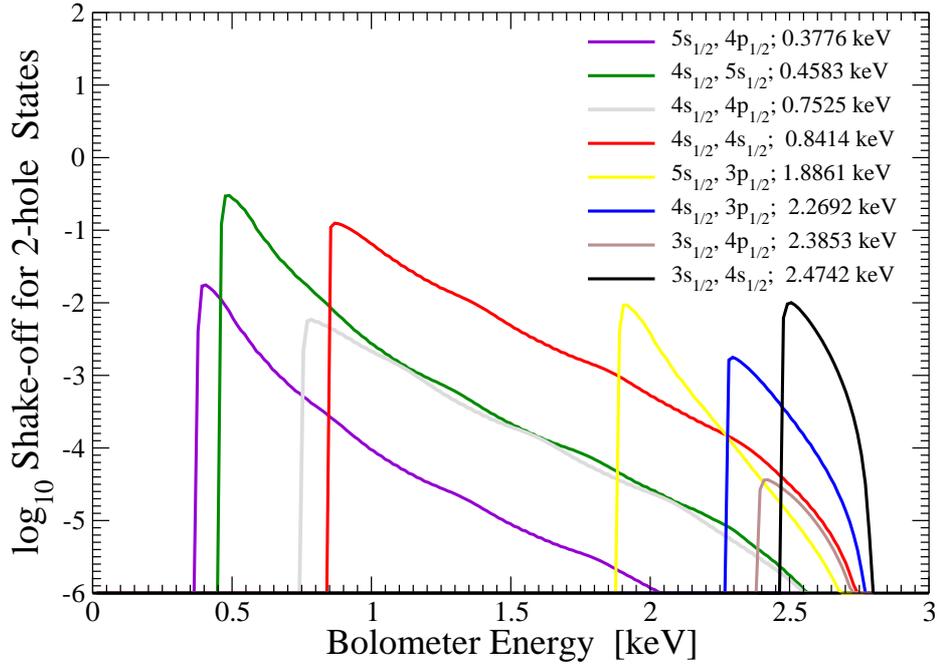}}
\caption{Shake-off contributions for different 2-hole excitations in Dy normalized for the experimental bolometer spectrum (see figure 10)  to the N1, $4s_{1/2}$ peak.  Increasing the energy $E_c$ of the bolometer spectrum  the Q-value = 2.8 keV is first used to excite the two hole state. So the shake-off contribution for the bolometer spectrum starts as function of $E_c$ with the 2-hole binding energy. Energy conservation yields an upper limit of Q = 2.8 keV for the bolometer spectrum. To integrate over the continuum energy of the shake-off electron (\ref{decay}) 
we divided the interval $<0.0\ ; \  2.8\ keV> $ into 417 mesh points.  From the left to the right with increasing bolometer 
energy $E_c$  the start of the different shake-off contributions are indicated in the figure. The energy difference between Q and $E_c$ is carried away by the neutrino, which can not contribute to the bolometer spectrum.}
\label{Fig4Greic}
\end{figure}
\begin{figure}
\centerline{\includegraphics[width=11cm,angle=-90]{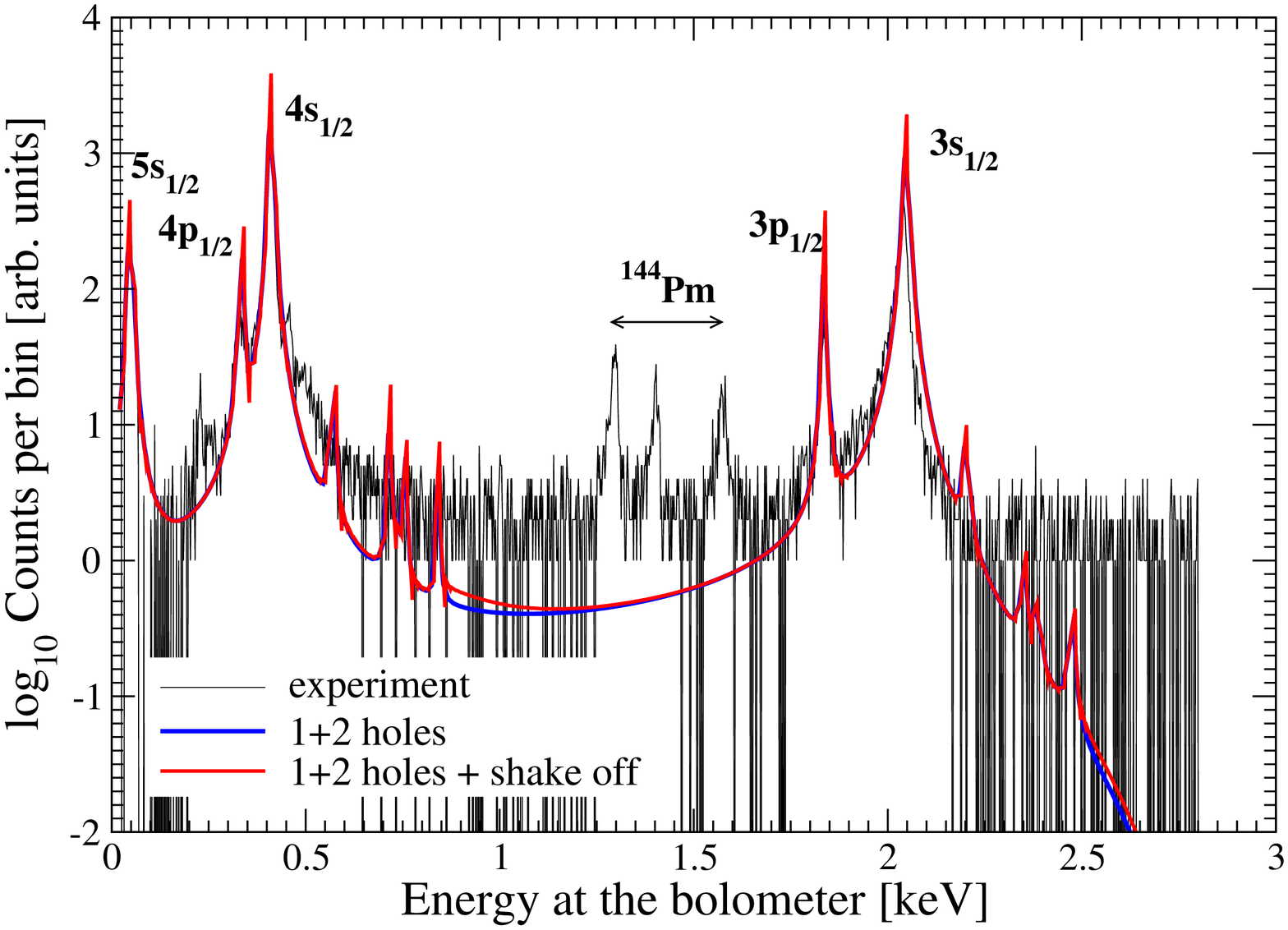}}
\caption{Experimental and theoretical results of the sum of the one- and two-hole deexcitations and the sum of the one-, two-hole and the shake-off deexcitation for the bolometer spectrum (\ref{decay}). The experimental data are from the ECHo collaboration 
\cite{Blaum} and \cite{Ra}. The two theoretical spectra are adjusted to  experiment at the  $N1, \ 4s_{1/2}$ peak. The nature of
the one hole states are indicated. The two-hole peaks are by about two orders of magnitudes smaller than the one hole peaks. The shake off contributions can hardly been seen in this scale. Some bins contain no experimental counts, thus the $log_{10}$ for these experimental values are minus infinity. To fit the 1931 experimental points for the bolometer energy of 0.0 to 2.8 keV, the theoretical spectrum of 200 mesh points had to be interpolated to the data points for this figure. Figure 3 contains the 200 original theoretical results without interpolation for the bolometer spectrum over $E_c$  between 0.0 and 2.8 keV. The interpolation is normally very good (compare figures 3 and 5) but can be difficult at some sharp minima and maxima.}
\label{Fig5Greic}
\end{figure}
The selfconsistent Coulomb field for the shake-off electron is in atomic units:
\begin{eqnarray}
V_{shake-off}(r) = \frac{-66}{r} +\sum_{k\ occupied \ e} g'_k \int \ d^3 \ r \cdot \frac{|\varphi_k(\vec{r}')|^2}{|\vec{r} - \vec{r}'|}
\label{pot2}
\end{eqnarray}
$g_k's$ are the number of bound electrons in the selfconsistent occupied orbits $|k>\ =\ |n,\ell,j>$  of Dy. 
To determine the potential for the shake-off electrons in Dy one needs the occupied selfconsistent Dirac-Hartree-Fock orbitals 
$P_k(r)$ and $Q_k(r)$. (As examples for P(r) see figure \ref{Fig2Greic}). 
\begin{eqnarray}
  \varphi_k(\vec{r})*r = (\ P_k(r) ; Q_k(r) ) \label{PQ}
\end{eqnarray} 
\begin{eqnarray} V_{shake-off} \approx -\frac{66}{r} + \frac{65}{r} \
\cdot [1 - exp(-a\cdot r)]\ [1/(length = a.u.)] ; \ \ a = 3.4  \label{pot3} \end{eqnarray} 
We adjust an analytic expression (\ref{pot3}) to the DHF potential (see figure \ref{Fig1Greic}). 
The selfconsistent Dy potential for the 66th. electron in the continuum is shown in figure \ref{Fig1Greic}  for Dy.
The wave functions are normalized with the WKB approximation to:
\begin{equation} \lim_{r \rightarrow \infty}\  P_k(r)\  = \ 2\cdot\sin(kr - \ell \cdot \frac{\pi}{2} -\eta \ln 2kr+ \Delta +\delta) 
\label{asymptotic} \end{equation} 
The normalization of the continuum Dirac wave function is treated in M. E. Rose "Relativistic electron theory " \cite{Rose} or Walter Greiner "Relativistic Quantum Mechanics" \cite{Greiner}. We follow here this recommendation of Perger and Karrighattam  on their page 394 \cite{Perger}. In the asymptotic expression  (\ref{asymptotic}) $\eta$ is the Sommerfeld parameter, $\Delta$  the Coulomb phase shift and $\delta$ takes into account deviations from a pure Coulomb potential. The electron energy in the continuum can be due to energy conservation not larger than the Q-value of 2.8 keV minus the excitation energy of the two hole state. So for shake-off with capture from 3s the energy of the electron in the continuum must be less than 0.758 keV (extreme non-relativistic). The most  important second hole is $4s1/2$ and thus for the two holes in 3s and 4s the integration over the shake-off electrons is restricted from zero to $2.800 -2.474 keV = 0.236 keV$. For capture from the two $4s1/2$ states the binding energy limits the integration in eq. (\ref{decay}) to an upperlimit of $2.800\  -\ 0.841\  = \ 1.959\  keV$. 
\\ \\
\newpage
\begin{eqnarray} \int_{r = 0 \ to \ \infty}  dr \cdot 2\cdot \sin(kr - \ell \cdot \frac{\pi}{2} -\eta \ln 2kr+ \Delta +\delta) \cdot \\ 2\cdot\sin(k'r - \ell \cdot \frac{\pi}{2} -\eta \ln 2kr+ \Delta +\delta) \approx 2\pi \cdot \delta(k-k'). \nonumber 
\label{knorm} \end{eqnarray} 
\\
The wave number is connected with the relativistic and non-relativistic energies by the equations:
\\
\begin{eqnarray} E^2_{rel} = c^2 \hbar^2 k^2 + m^2 c^4 \ \rightarrow \ c^2 k^2 + c^4\ \ (in\  atomic \ units);\ \  E_{n-rel} = \frac{1}{2}k^2; \hspace{1.5cm} \nonumber  \\ 
 k = \alpha \sqrt{E_{n-rel}(E_{n-rel} +  2\cdot c^2)}; \ \ \ with \ c \ = \ 1/\alpha \ = \ 137.035999\  in\ [atomic\  units].
\label{wavenumber} \end{eqnarray} 
The transformation from the asymptotic wave number normalization $P_k(r)$ to the asymptotic energy delta function normalization is:
\begin{eqnarray} P_E(r) = P_k(r) \cdot \sqrt{\frac{1}{2\pi} \cdot  \frac{1}{c} \cdot \frac{ \sqrt{k^2+c^2}}{k}}\ \approx P_k(r)\cdot \frac{1}{\sqrt{2 \cdot \pi\ \cdot k}} \hspace{1cm} \\ \nonumber
\delta(E[Hartree])\  =\  \delta(36.74932386 \cdot E[keV])\  = \ 0.027121138506 \cdot \delta(E[keV])
\label{energy} 
\end{eqnarray} 
\\
For shake-off one has to calculate the overlap of the bound Ho electron orbitals $|p>$ with the continuum wave functions in Dy $|q'>$ i.e.
$<q'_{>0}|p_{<F}>$ (\ref{con}) (see figure \ref{Fig2Greic}). For this one expands the configuration of Ho after capture of the bound electron $|b,Ho>$ with now the same number of protons as Dy (- but not a Dy eigenstate -) into the complete set of configurations $|D, Dy>$  in Dy including the continuum.
\newpage
\begin{eqnarray} 
|P, \ (b)^{-1}, \ Ho> = \sum_{D \neq P \ , bound} a_{D} \cdot |D,\ Dy> + \int_{0\ to\ \infty} dE' \cdot a(D,\ E') |D, \ E',Dy> \hspace{14 cm} \nonumber \\
a_{D} = <D ,Dy|(b)^{-1}|P, \ (b)^{-1}, Ho>, \hspace{18cm} \nonumber \\
a(D, \ E'') = <D, E'',Dy|P,(b)^{-1}, Ho> =  \hspace{18cm} \nonumber \\
\int_{0 \ to\ \infty} \cdot dE' \cdot a(D, E') \cdot <D, E'',Dy|D, E',Dy> \hspace{3cm} (22) \hspace{14cm}
\label{expansion}
\end{eqnarray} 
The probability forming a specific hole state $|k',Dy>$ in Dy in a bound orbit after capture of the electron $|b,Ho>$ is proportional to $|<k',Dy|P,\ (b)^{-1},Ho>|^2$ and and for the continuum $|E'',Dy>$ to $|<E'',Dy|P,\  (b)^{-1}, Ho>|^2$.
\section{Shake-off Probabilities.}
The overlap between the bound Ho states $ns_{1/2}$ and $np_{1/2}$ with $n \ge 3$ and the continuum wave functions in Dy squared gives the probability for shake-off. We restrict this work to s-wave shake-off, which is expeted to be the largest contribution.  The s-wave shake-off  is given by the overlap $<n \ge3, s_{1/2}, Ho|E,\ s, Dy>$ squared. In the spectrum with 1-hole, 2-hole and shake-off in figure \ref{Fig3Greic}  and \ref{Fig5Greic} the shake-off contribution is hardly  visible. Here and also in ref. \cite{de2} only the decay width of the 2-hole states are included. Three hole states can be neglected \cite{Fae3hole}. The electron in the continuum has an  escape width, which is not included. 
Figure \ref{Fig3Greic} shows the logarithmic spectrum of the 1-hole, the 2-hole and the s-wave shake-off contributions. The shake-off contributions are calculated for the different 2-hole states listed in table \ref{2-binding}. The two-hole spectrum is about two orders of magnitude smaller than the one-hole states. The shake-off spectrum can hardly  be seen on this scale in the total spectrum. Compared to the one-hole peaks it is at least two orders smaller. The integration over the continuum electron energy (\ref{decay}) is  done by the Bode method using 417 mesh points.  Shake-off is proportional to the square of the overlap $<Ho-bound|Dy-continuum> $. The 2-hole states contributing to s-wave shake-off are listed in table \ref{2-binding}.  The shake-off  contributions of the 2-hole states as function of the bolometer energy $E_c$ is starting from the 2-hole binding energy up (see table \ref{2-binding}). The two main contributions originate from 4s1/2, 5s1/2 starting at 0.4583 keV and from 4s1/2, 4s1/2 starting at 0.8414 keV. The $log_{10}$ contributions from 3s1/2, 4p1/2 starting at 2.33853 keV (see table \ref{2-binding}) are extremely small. The energy difference $Q - E_c$ is carried away by the neutrino and and does not show in the bolometer. 
\section{Summary}
In this contribution to the memorial volume for Walter Greiner I summarized our work on the determination of the electron neutrino mass by electron capture \cite{ Fae3,Fae4,Fae7,Fae8,Fae3hole, Fae1-2hole} in $^{163}_{67}Ho$ to $^{163}_{66}Dy$. The energy of the $Q-value = 2.80\pm0.08\ keV$ isdistributed to the excitation of Dy and the emission of a neutrino. The recoil of $^{163}Ho$ can be neglected. The deexcitation as X-rays and Auger electrons of Dy can be measured in a bolometer. The upper end of the spectrum requires the minimum energy of the neutrino. Thus the difference of Q minus this upper limit is the rest mass of the electron neutrino. 
For the excitation of Dy* we alowed one-hole excitations\cite{Fae3}, one- and two-hole excitations \cite{Fae1-2hole} (shake-up),  three-hole excitations \cite{Fae3hole} (shake-up) and the ecitation into the electron continuum \cite{Fae4} in Dy* (Shake-off). 
The excitation of Dy* is calculated by the overlap squared of Ho* with a hole in the orbit, from which the electron is captured e. g. $3s_{1/2}, 3p_{1/2}, 4s_{1/2}, ...$, and the excited configurations in Dy*.  Equations $(\ref{overlapA})$ and $(\ref{VataiA})$ show, that a small change of this overlap of about 10 \% can produce a change of the shake-up or shake-off by about a factor 100. Eqs. $(\ref{overlapA})$ and $(\ref{VataiA})$ serve as lever for large shake-up and shake-off effects. Inspite of this lever the estimate of $(\ref{VataiA})$ shows, that shake-up must be less the 0.4 \% of the probability for the one-hole excitations. 
The 1-hole and 2-hole electron configurations play the importat role for the Dy* excitation. The three hole excitations and the excitation of a larger number of holes can be neglected \cite{Fae3hole}. The total 2-hole (shake-up) and continuum (shake-off) excitation probability 
is given in the sudden plus the Vatai \cite{Vatai1, Vatai2} approximation by unity minus the overlap squared between Ho* minus the captured electron and Dy* configurations  with the number of electrons in the exponent with the same  caputure orbit quantum numbers in Dy as in Ho (\ref{overlapA}) and (\ref{VataiA}).
\begin{eqnarray}  (1.0 - <Ho^{*}, n, \ell,j|Dy,n,\ell,j>^{2(2j+1)})  \label{overlap2} \end{eqnarray}
These Ho-Dy overlaps have in selfconsistent relativistic Dirac-Hartree-Fock values of about 0.999 and even closer to unity  (see table \ref{over}). This yields the above already mentioned total 2-hole plus shake-off probability of about $0.4 \ \%$. This estimate is only very approximate. (The final results in section 2 to 5 are all calculated without the Vatai approximation.)Important is the result, that a small uncertainty for these electron overlaps between Ho and Dy produces a large increase for the 2-hole and the shake-off probability, which meight be  about two orders of magnitude for shake-off. Thus it is not permitted to use very approximate wave functions 
\cite{de2} for Ho* and Dy*. We describe the bound states in Ho and Dy by the  Dirac-Hartree-Fock approach \cite{Ankudinov, Grant, Desclaux} even including  occupations in Dy due to different  hole states. The s-wave continuum wave functions in Dy are determined with the Dirac equation in the selfconsistent potential \cite{Salvat}.(See figures: (\ref{Fig1Greic}) and (\ref{Fig2Greic})).
The energy of the capture orbit in Ho and the excitations of Dy* including also the continuum states are limited by energy conservation to the Q-value minus the xcited  one-hole states in Ho* and Q minus the ecitation   energies in Dy*. E. g. 
the $ 3s_{1/2}, \ 4s_{1/2}$ 2-hole state limits the upper bound of the continuum energy contributions to 2.8 - 2.4742 = 0.3258 keV. So this contribution is very small. One of the two  main contributions comes from the 2-hole state $4s_{1/2},\  4s_{1/2}$ with a binding energy of 0.841 keV and thus an upper limit of the continuum energy of 2.8 -  0.841 = 1.959 keV. A second large contribution is built on the 2-hole state $ 4s_{1/2},\  5s_{1/2}$ with the binding energy 0.4398 keV. Thus the upper limit of the shake-off contributions in the continuum integration (\ref{decay}) is the 2.8 - 0.4398 = 2.3602 keV for the integral over the shake-off continuum electron. 
\newline 
We prepared two different computer programs  both calculating the s-wave shake-off to test the two codes against each other. All the calculations are done in double precision and for the critical integrations we use parallel the Trapez, the Kepler-Simpson, the Bode and the Weddle rules to test the accuracy. The numbers given are the ones from the Bode rule. (Trapez is not reliable enough.) The contributions  from the shake-off process are small (see figures \ref{Fig3Greic}, \ref{Fig4Greic}, \ref{Fig5Greic}). The widths for the shake-off states include only the values from the 2-hole excitations as in \cite{de2}. In reality one has to include the escape width of the electron in the continuum, which could perhaps even be larger than the contribution of the 2-hole states.
\newline
In summary this work shows, that one has not to worry about the three hole excitations and the shake-off process 
in $^{163}Dy*$ in the determination of the neutrino mass from electron capture in  $^{163}Ho$. 
\newline 
The remaining discrepancies between theory and experiment, e. g. the slope above the 1-hole state $4s_{1/2} \ (N1)$, are probably due to configuration mixing not included  here. 
Finally we want to stress, that the accuracy needed to extract the neutrino mass  can not be obtained by theoretical calculation alone. One must fit simultaneously the neutrino mass, the Q-value, the excitation energy of the highest resonance, their width and their strength at the upper end of the spectrum to extremely accurate data. For this the present background of the experimental 
spectrum has to be drastically reduced. The measurement must probably be done in an underground 
lab to protect against cosmic rays. This reduces the background as a first measurement in the Modane underground lab in the 
Fr\'{e}jus tunnel has shown \cite{Loredana}.  If the background can be reduced by one to two orders, it should be enough to adjust only the two parameters, the Q-value and the neutrinos mass, to the upper end of the spectrum for the determination of the electron neutrino mass. 
\section{Acknowlegement} 
 I want to thank Christian Enss, Loredana Gastaldo and Fedor \v{S}imkovic for their collaboration in earlier works
\cite{Fae3, Fae7, Fae8, Fae3hole, Fae1-2hole}  on this topic. 
\bibliographystyle{ws-rv-van}

\end{document}